\documentstyle[aps,preprint]{revtex}

\newcommand{\Ket}[1]{\left\vert#1\right\rangle}
\newcommand{\BraKet}[2]{\left\langle#1\right\vert\left.#2\right\rangle}
\newcommand{\KetBra}[2]{\left\vert#1\right\rangle\left\langle#2\right\vert}
\newcommand{\MatrixEl}[3]{\left\langle#1\right\vert #2 \left\vert#3\right\rangle}
\newcommand{\MeanValue}[1]{\left\langle#1\right\rangle}

\newcommand{\bdmreal}    {\mbox{{\bf R}\hspace{-3.7 mm}{\sf I}\hspace{3mm}}}
\newcommand{\bdmnatu}    {\mbox{{\bf N}\hspace{-3.7 mm}{\sf I}\hspace{3mm}}}

\begin{document} \draft \textheight=9in \textwidth=6.5in \draft %PRA style

\title{Measuring the mean value of vibrational observables in trapped ion systems}
\author{B. Militello, A. Messina%\footnote{e-mail: messina@fisica.unipa.it}
        , A. Napoli\\
        \emph{INFM, MURST}\\
        \emph{and Dipartimento di Scienze Fisiche ed Astronomiche} \\
        \emph{via Archirafi 36, 90123 Palermo (ITALY)}
        }
%\date{\today}
\maketitle

\begin{abstract}
The theoretical foundations of a new general approach to the measurement problem of vibrational observables in trapped ion systems is reported. The method rests upon the introduction of a simple vibronic coupling structure appropriately conceived to link the internal ionic state measurement outcomes to the mean value of a motional variable of interest. Some applications are provided and discussed in detail, bringing to light the feasibility and the wide potentiality of the proposal.
\end{abstract}


\pagebreak

\section{Introduction}

The question of how measuring the probability distribution of the eigenvalues of an observable $\hat{A}$ in a given quantum state often exhibits, at least, a high level of practical complexity. Sometimes, however, all what we want to know about the physical quantity under scrutiny is just its expectation value. To extract this information is of course possible if the probability for each eigenvalue of $\hat{A}$ in the given state is known, even if, in principle, such a detailed knowledge goes beyond the specific need.

In the last few years advancements in ionic trapping and cooling techniques have provided the possibility of carefully controlling the dynamics of a  two-level object coupled to a bosonic system\cite{nist,Vogel-Rass}.
In a Paul microtrap, for example, a suitable configuration of inhomogeneous and alternating electromagnetic fields force the centre of mass of a confined two-level ion to behave as a quantum harmonic oscillator\cite{Ghosh,Toschek}.
When appropriate laser beams act upon the system, a vibronic coupling is switched on.
Thus one may efficiently manipulate the trapped ion quantum state in order to realise, for example, applications in the quantum logic context\cite{q-logic} and non-classical states\cite{non-classic}.

Proposals of measurement processes aimed at revealing the quantum dynamics of such systems have been reported.

In 1997 L.G.Lutterbach and L.Davidovich have presented a proposal to extract all possible information about the vibrational quantum state of a trapped ion centre of mass\cite{Davidovich}. Such a knowledge is reached via the reconstruction of the Wigner distribution, $W(\alpha)$, relative to the ion centre of mass degrees of freedom ($\alpha=x+i p$, $x$ and $p$ being the \emph{dimensionless} harmonic oscillator position and momentum respectively).
It is well known that the value of the Wigner function in the phase space origin coincides with the mean value of the parity  operator in the relative quantum state\cite{ParityOp}.
Moreover, subjecting the system to some suitable external forces whose action may be described by displacements operators, one may induce a phase space translation of the physical system and hence a phase space translation of the Wigner distribution, passing from $W(\alpha)$ to $\overline{W}(\alpha)\; :\;\overline{W}(\alpha-\alpha_0)=W(\alpha)$, $\alpha_0$ being related to the external action.
Exploiting the two features just recalled one has, in principle, a recipe to measure $W(\alpha)$ at any point.
Indeed the mean value of the parity operator evaluated in the system Wigner function translated by $\alpha_0$, is just the value of the original Wigner function in the point $\alpha_0$.

If the interest is pointed in the individuation of the vibrational quantum energy probability distribution only, one can turn to quantum nondemolition (QND) measurements as already proposed in two-level trapped ion systems both in one-dimensional\cite{Vogel-QND,QND-group} and two-dimensional\cite{Sabrina-QND} contexts.
Generally speaking, in QND measurements a \emph{signal} observable is indirectly measured observing a \emph{meter} quantity which has been suitably modified via an appropriate \emph{signal-meter interaction}\cite{Braginsky,Walls}. If the signal is conserved both by the free hamiltonian and by the signal-meter interaction, then the signal observable may be repeatedly measured getting rid of disturbances due to the \lq\lq free time evolution\rq\rq (between two measurements) or to the \lq\lq back-action\rq\rq (during the signal-meter interaction).

In the trapped ion context, QND schemes have been applied taking the vibrational energy as the signal and the atomic populations as the meter\cite{Vogel-QND,QND-group,Sabrina-QND}. The method is based upon the ability of implementing a suitable vibrational energy conserving vibronic coupling that resonantly drives weak atomic transitions of the atom beyond the Lamb-Dicke limit (signal-meter interaction), and upon the possibility of monitoring such transitions (meter observations). For example in ref\cite{QND-group} it has been proposed a quantum nondemolition measurement scheme of the vibrational population of a trapped ion, by means of Raman pulses inducing resonant electronic transitions. The fundamental point of this method is the dependence of the Rabi oscillations between the two electronic levels on the energy quantum number.

The two protocols just recalled allow to realise measurements of quantum observables or even reconstructions of the entire dynamical configuration of the system, but both are generally very complicated to be executed. On the other hand, in many cases the knowledge of the mean value of some suitable observable is enough to reveal the phenomena we are interested in.

In this paper we present a new general approach to the measurement of the mean value of a vibrational dynamical variable in the trapped ion context. The method, its reliability and its feasibility are described in detail in the second section. In the subsequent section some possible applications are described bringing to light the effectiveness and the wide potentiality of our proposal.

\section{The idea and the method}

The physical system on which we focus consists in a two-level ion confined into a three-dimensional Paul trap\cite{nist,Vogel-Rass,Ghosh,Toschek}.
In such a trap the motion of the ion centre of mass is confined in an effective harmonic potential which, under appropriate conditions, may be rendered isotropic\cite{Toschek}, so that the free dynamics of our system is governed by the free hamiltonian
\begin{equation} \label{unperturbed}
  \hat{H}_0=\hbar\omega_{T}\sum_{j=x,y,z}\hat{a}_j^{\dag}\hat{a}_j
           +\hbar\omega_{A} \hat{\sigma}_{z}
\end{equation}
Here $\omega_T$ is the frequency of the trap, $\hat{a}_i(\hat{a}_i^{\dag})$ with $i=x,y,z$ are the annihilation (creation) operators of the centre of mass oscillatory motion, $\omega_A$ is the atomic transition frequency and $\hat{\sigma}_z$ the $z$ component of the pseudospin operator $\stackrel{\rightarrow}{\sigma}\equiv(\hat{\sigma}_x,\hat{\sigma}_y,\hat{\sigma}_z)$ associated to the internal ionic dynamics. Here $\hat{\sigma}_x=\KetBra{+}{-}+\KetBra{-}{+}$, $\hat{\sigma}_y=i(\KetBra{+}{-}-\KetBra{-}{+})$, $\hat{\sigma}_z=\KetBra{+}{+}-\KetBra{-}{-}$ and $\Ket{\pm}$ are the two effective atomic levels such that $\hat{\sigma}_z\Ket{\pm}=\pm\Ket{\pm}$.

In the experiments currently performed on trapped ion systems, the probability of finding the ion in its \emph{ground} ($\Ket{-}$) or \emph{excited} ($\Ket{+}$) electronic state may be measured using quantum jumps techniques with almost $100\%$ efficiency\cite{exp-values-nist}. The measurement of bosonic dynamical variables of interest is in practice systematically
traced back to atomic population observations.

In this paper we sketch an experimental scheme that, in principle, allows to measure the mean value of vibrational dynamical variables.
The key idea of our proposal relies on the ability of implementing an appropriate vibronic coupling by which we succeed in making the atomic populations after the interaction dependent on the mean value of the vibrational observable in the initial state, that is  before the interaction.
In other words, starting from a dynamical configuration, we induce a time evolution able to correlate electronic populations to the mean value of the observable in the initial vibrational state.

Let $\hat{A}$ be the \emph{dimensionless} vibrational (or bosonic)
observable under scrutiny, given in the Schr\"odinger picture, and
assume it is time independent. Suppose now to subject the system
to a suitable laser configuration responsible for realising a
vibronic coupling of strength $\gamma$ that, in the interaction
picture relative to $\hat{H}_0$, has the form
\begin{equation} \label{interaction}
  \hat{H}^{(I)}_I=\hbar\gamma\hat{A}\hat{\sigma}_{x}
\end{equation}

We will show that the hamiltonian given in eq.(\ref{interaction}) is responsible for inducing atomic transitions \emph{modulated} by the mean value of $\hat{A}$, $<\hat{A}>$, evaluated in the initial state.

In the simple case of a trapped ion prepared in the state
\begin{equation} \label{alfa-init-cond-z}
  \Ket{\psi(0)}=\Ket{\alpha}\Ket{-}
\end{equation}
with $\Ket{\alpha}$ such that $\hat{A}\Ket{\alpha}=\alpha\Ket{\alpha}$,
the interaction picture quantum dynamics under the action of the hamiltonian given in eq.(\ref{interaction}) is
\begin{equation} \label{eigenst-time-evol}
  \Ket{\psi^{(I)}(t)}=   \cos(\gamma\alpha t)\Ket{\alpha}\Ket{-}
                -i \sin(\gamma\alpha t)\Ket{\alpha}\Ket{+}
\end{equation}

Looking at this time evolution we note that the Rabi oscillation frequency is proportional to $\alpha$. Hence, as given in advance, the flipping rate from $\Ket{-}$ to $\Ket{+}$ is just proportional to the mean value of $\hat{A}$ in the initial vibrational state: $\MatrixEl{\psi(0)}{\hat{A}}{\psi(0)}\equiv\MatrixEl{\psi_{vibr}}{\hat{A}}{\psi_{vibr}}=\alpha$.

Obviously in the more general case of an arbitrary bosonic initial state
\begin{equation} \label{init-cond-z}
  \Ket{\psi(0)}=\Ket{\psi_{vibr}}\Ket{-}
\end{equation}
the analysis is not so simple. Nevertheless it is always possible to establish a correlation between electronic transitions and the initial mean value of $\hat{A}$. To this purpose, subject the ion to a $\frac{\pi}{2}$-pulse relative to the interaction of the form $\hbar\delta\sigma_x$ ($\delta$ being a real number representing the strength of the interaction), easy to implement with a laser tuned on the electronic transition frequency ($\omega_A$)\cite{nist,Vogel-Rass}. The effect on the system is the passage to the state
\begin{equation} \label{init-cond-y}
  \Ket{\overline{\psi}(0)}=\Ket{\psi_{vibr}}\Ket{-}_y
\end{equation}
where $\Ket{-}_y$ is an eigenstate of the operator $\hat{\sigma}_{y}$ and $\Ket{\psi_{vibr}}$ is the same generic vibrational state appearing in eq.(\ref{init-cond-z}).

Calculate now the time evolution of the $\hat{\sigma}_z$ mean value under the action of the hamiltonian in eq.(\ref{interaction}). Since $[\hat{H}_0,\hat{\sigma}_z]=0$ the interaction picture operator $\hat{\sigma}_z^{(I)}$ coincides with the Schr\"odinger picture operator $\hat{\sigma}_z$, and hence
\begin{eqnarray}
  \nonumber
  \MeanValue{\hat{\sigma}_z(t)}&\equiv&
    \MatrixEl{\psi^{(I)}(t)}{\hat{\sigma}_z^{(I)}}{\psi^{(I)}(t)}=     \MatrixEl{\psi^{(I)}(t)}{\hat{\sigma}_z}{\psi^{(I)}(t)}=\\
    &=&\MatrixEl{\overline{\psi}(0)}{\hat{U}_A^{\dag}(t)\hat{\sigma}_z
                 \hat{U}_A(t)}{\overline{\psi}(0)}
\end{eqnarray}
with
\begin{equation}
  \hat{U}_A(t) = e^{-i\frac{\hat{H}^{(I)}_I}{\hbar}t}
\end{equation}
the interaction picture time evolution operator.

Moreover, being
\begin{equation}\label{canonical-trasf}
  \hat{U}_A^{\dag}(t) \; \hat{\sigma}_z \; \hat{U}_A(t) =
               \cos(2\gamma\hat{A}t) \hat{\sigma}_z +
               \sin(2\gamma\hat{A}t) \hat{\sigma}_y
\end{equation}
one immediately obtains, for the state in eq.(\ref{init-cond-y}),
\begin{eqnarray}\label{el-bos}
  \MeanValue{\hat{\sigma}_z(t)}=
  -\MatrixEl{\psi_{vibr}}{\sin(2\gamma\hat{A} t)}{\psi_{vibr}}
\end{eqnarray}

For the sake of completeness, consider the case of an operator having a non-degenerate spectrum being partly continuous and partly discrete. Let $\Omega_D=\{\alpha_n\}$ be the set of all the discrete eigenvalues and $\Omega_C$ the set corresponding to the continuous spectrum of $\hat{A}$.

The vibrational state $\Ket{\psi_{vibr}}$ can be expanded in terms of the states $\Ket{\alpha}$ as follows:
\begin{eqnarray}\label{expansion}
  \nonumber
  \Ket{\psi_{vibr}}=\sum_{\alpha\epsilon\Omega_D}
                          \BraKet{\alpha}{\psi_{vibr}}\Ket{\alpha}d\alpha+\\
                    \int_{\alpha\epsilon\Omega_C}
                          \BraKet{\alpha}{\psi_{vibr}}\Ket{\alpha}d\alpha
\end{eqnarray}

Starting from this expansion, the mean value of $\sin(2\gamma\hat{A} t)$ appearing in eq.(\ref{el-bos}), may be cast in the form

\begin{eqnarray}
  \nonumber
  \MatrixEl{\psi_{vibr}}{\sin(2\gamma\hat{A}t)}{\psi_{vibr}}=
     \sum_{\alpha\epsilon\Omega_D}
     |\BraKet{\alpha}{\psi_{vibr}}|^2\sin(2\gamma\alpha t)d\alpha+\\
     \int_{\alpha\epsilon\Omega_C}
     |\BraKet{\alpha}{\psi_{vibr}}|^2\sin(2\gamma\alpha t)d\alpha
\end{eqnarray}

In many physical situations of interest, it is reasonable to assume that $|\BraKet{\alpha}{\psi_{vibr}}|^2$ may be neglected when $\alpha$ is outside the interval
$[-\alpha_{max},\alpha_{max}]$, $\alpha_{max}$ being an appropriate positive real number (we shall refer to this assumption as \emph{finite effective spectrum hypothesis}). At this point we wish to emphasise that the possibility of such an assumption  is not necessarily related to an \textit{a priori} knowledge of $\vert\psi_{vibr}\rangle$.
Consider for example the case of $\hat{A}=\hat{X}$ ($\hat{X}$ being the $x$-component of the ion centre of mass position operator given in the Schr\"odinger picture). For such an operator
it results $\Omega_D=\emptyset$ and $\Omega_C=\bdmreal$. Nevertheless, due to the ion confinement, one may be confident in the fact that all the effectively involved values of the
observable $\hat{X}$ must be compatible with the trap dimensions.

As another example consider the bosonic excitation number operator $\hat{n}_i$ (with $i=x,y,z$). One has $\Omega_C=\emptyset$ and, in principle, $\Omega_D=\bdmnatu$. Nevertheless, again because of the confinement, the probability of finding the charged trapped particle with $n_i$ beyond an appropriate maximum value, may be reasonably assumed as being negligible.

It is worth noting that the $z$ component of the orbital angular momentum $\hat{L}_z$, is canonically equivalent to the operator $\hat{n}_x-\hat{n}_y$. Thus, bearing in mind the previous considerations regarding $\hat{n}_i$, we legitimately claim that also $\hat{L}_z$ satisfies the finite spectrum hypothesis in the experimental situation under scrutiny. We remark that this property holds for any component of the orbital angular momentum due to the spherical symmetry of the problem. Analogous considerations are applicable to the correlation operators as, for example, $\hat{C}_{xy}\equiv(\hat{a}_x^{\dag}\hat{a}_y+\hat{a}_y^{\dag}\hat{a}_x)$, reaching the same conclusions. Indeed, like $\hat{L}_z$, also the operator $\hat{C}_{xy}$ is canonically equivalent to $\hat{n}_x-\hat{n}_y$, being $e^{i\frac{\pi}{4}\hat{L}_z}\hat{C}_{xy}e^{-i\frac{\pi}{4}\hat{L}_z}=e^{i\frac{\pi}{4}\hat{C}_{xy}}\hat{L}_{z}e^{-i\frac{\pi}{4}\hat{C}_{xy}}=\hat{n}_x-\hat{n}_y$.

Under the \emph{finite spectrum hypothesis} we may control both the coupling parameter $\gamma$ and the interaction duration $t$ in such a way that $\sin(2\gamma\alpha t)\simeq 2\gamma\alpha t$ for any $\alpha$ inside $[-\alpha_{max},\alpha_{max}]$.
(In the following we shall refer to the interval of real values for which $\sin x \simeq x$ as the \emph{linearisability zone}.)

Such an assumption makes it legitimate linearising the mean value of the operator $\sin(2\gamma\hat{A}t)$, which means, reducing it to the mean value of $2\gamma\hat{A}t$. Indeed, since the only relevant contributions to the mean value calculations are the ones relative to $\alpha$ belonging to $\Omega_C'=\Omega_C\cap [-\alpha_{max},\alpha_{max}]$ or $\Omega_D'=\Omega_D\cap [-\alpha_{max},\alpha_{max}]$, one deduces

\begin{eqnarray}\label{approx-sin-1}
  \nonumber
  \MatrixEl{\psi_{vibr}}{\sin(2\gamma\hat{A}t)}{\psi_{vibr}}
    &=&
    \sum_{\alpha\epsilon\Omega_D}
          |\BraKet{\alpha}{\psi_{vibr}}|^2\sin(2\gamma\alpha t)d\alpha\\
  \nonumber
    &+&\int_{\alpha\epsilon\Omega_C}
          |\BraKet{\alpha}{\psi_{vibr}}|^2\sin(2\gamma\alpha t)d\alpha\simeq\\
  \nonumber
    &\simeq&
    \sum_{\alpha\epsilon\Omega_D'}
          |\BraKet{\alpha}{\psi_{vibr}}|^2\sin(2\gamma\alpha t)d\alpha+\\
  %\nonumber
    &+&\int_{\alpha\epsilon\Omega_C'}
          |\BraKet{\alpha}{\psi_{vibr}}|^2\sin(2\gamma\alpha t)d\alpha\simeq
\end{eqnarray}

Since the calculation is developed in the linearizability zone, it
is possible to linearise the sinusoidal function before coming
back to the original dominion of the variable $\alpha$:
$\Omega_D\cup\Omega_C$. Hence it follows that

\begin{eqnarray}\label{approx-sin-2}
  \nonumber
  \MatrixEl{\psi_{vibr}}{\sin(2\gamma\hat{A}t)}{\psi_{vibr}}
   &\simeq&
    \sum_{\alpha\epsilon\Omega_D'}
          |\BraKet{\alpha}{\psi_{vibr}}|^2 2\gamma\alpha td\alpha+\\
  \nonumber
    &+&
    \int_{\alpha\epsilon\Omega_C'}
          |\BraKet{\alpha}{\psi_{vibr}}|^2 2\gamma\alpha td\alpha\simeq\\
  \nonumber
    &\simeq&
    \sum_{\alpha\epsilon\Omega_D}
          |\BraKet{\alpha}{\psi_{vibr}}|^2 2\gamma\alpha td\alpha+\\
  \nonumber
    &+&
    \int_{\alpha\epsilon\Omega_C}
          |\BraKet{\alpha}{\psi_{vibr}}|^2 2\gamma\alpha td\alpha
    \equiv\\
    &\equiv&
    \MatrixEl{\psi_{vibr}}{2\gamma\hat{A}t}{\psi_{vibr}}
\end{eqnarray}

After such a discussion we are ready to deduce the main result of this paper. Assuming that $\hat{A}$ has an appropriately bounded spectrum and accordingly controlling the pulse area $\gamma t$, we may take advantage of eq.(\ref{approx-sin-2}) into eq.(\ref{el-bos}) obtaining
\begin{equation}\label{Mean-Sigma-A}
  \MeanValue{\hat{A}(t=0)}\equiv\MatrixEl{\psi_{vibr}}{\hat{A}}{\psi_{vibr}}\simeq
  -\frac{1}{2\gamma t}\MeanValue{\hat{\sigma}_z(t)}
\end{equation}

Eq.(\ref{Mean-Sigma-A}) relates in a transparent way the mean value of the vibrational observable $\hat{A}$ before the interaction in eq.(\ref{interaction}) (that is in the initial vibrational state), with the mean value of the $z$-pseudospin component, $\MeanValue{\hat{\sigma}_z(t)}$, measured immediately after this interaction.

We again emphasise that the applicability of
eq.(\ref{Mean-Sigma-A}) requires, among other conditions, the
narrowness of the interval containing the eigenvalues of $\hat{A}$
appearing in the expansion of the initial vibrational condition.
The importance of this point may be fully appreciated considering
that the identification of the mean value of
$\sin(2\gamma\hat{A}t)$ with the mean value of $2\gamma\hat{A}t$
amounts at  neglecting the mean values of all the nonlinear terms
appearing in the series defining the operator
$\sin(2\gamma\hat{A}t)$, that is at assuming
\begin{equation}\label{desc}
  \sum_{k=1}^{\infty}
    \frac{\MeanValue{\hat{A}^{2k+1}} \left(2\gamma t\right)^{2k+1}}{({2k+1})!}
    <<2\gamma t \MeanValue{\hat{A}}
\end{equation}
Indeed, the finite spectrum hypothesis allows to claim that $|\MeanValue{\hat{A}^n}|\leq \alpha_{max}^n$ legitimating relation (\ref{desc}) when $\gamma\alpha_{max}t<<1$.

For the sake of clarity we summarise the steps of the process we propose for measuring the mean value of vibrational observables in the state $\Ket{\psi_{vibr}}$:

a) Estimate an upper limit, $\alpha_{max}$, for the modulus of the
bosonic eigenvalues of $\hat{A}$ involved in $\Ket{\psi_{vibr}}$.
As previously stressed, such an estimation does not require the
\textit{a priori} knowledge of $\Ket{\psi_{vibr}}$.

b) Convert the initial condition given by eq.(5) into the state $\Ket{\psi} = \Ket{\psi_{vibr}}\Ket{-}_y$ by the action of a $\frac{\pi}{2}$-pulse relative to the interaction $\hbar\delta\hat{\sigma}_x$.

c) Subject the system to a pulse related to the interaction $\hat{H}_I^{(I)}=\gamma\hat{A}\hat{\sigma}_x$. The intensity and the duration of the pulse must be chosen in such a way that $2\gamma \alpha_{max} t$ lies in the \emph{linearizability zone}.

d) Measure the atomic populations, calculate $\MeanValue{\hat{\sigma}_z}$ and hence deduce $\MatrixEl{\psi_{vibr}}{\hat{A}}{\psi_{vibr}}$ from eq.(\ref{Mean-Sigma-A}).\\

We conclude this section giving some orders of magnitude for the parameters involved in the measurement process. To this end we consider for example $\hat{A}=\hat{n}_i$.
According to ref.\cite{nist,exp-values-nist}, $n_i=20$ is surely a good estimation for a maximum value of the bosonic excitations, in the sense that the experiments up to now performed generally involve just few phonons.

About the coupling strength $\gamma$, we may refer both to the Innsbruck group experiments (with $Ca^+$ ions), where typical $\gamma$ values are of the order of magnitude of $10KHz$\cite{exp-values-Inn}, or to Boulder group experiments (with $Be^+$ ions), where typical $\gamma$ values are of hundreds of $KHz$\cite{exp-values-nist}. Although the latter values are higher than the former ones, we guess it should be possible to obtain smaller values even with $Be^+$ ion experiments just decreasing the laser intensities.

Assuming therefore $\gamma\sim 10KHz$ and $[-0.4,0.4]$ as the
linearizability zone, in the case of $\hat{A}=\hat{n}_i$, one
estimates an interaction time $t\sim1\mu s$ which is in the grasp
of experimentalists\cite{nist}. Indeed we emphasise that the brief
duration of the vibronic interaction is compatible with the laser
\lq\lq turning-on\rq\rq-\lq\lq turning-off\rq\rq and also makes
our method reasonably immune from decoherence effects.

\section{Applications and conclusive remarks}

A fundamental underlying requirement of the method presented in
the previous section is the implementation of the hamiltonian
given by eq.(\ref{interaction}) relative to the specific bosonic
observable one wishes to measure. Fortunately the trapped ion
context allows to tailor practically at will vibronic couplings
possessing a desired form. By changing indeed the lasers
configuration and parameters, a huge class of effective
Hamiltonian operators may be engineered.

In this section we focus our attention on some vibrational
observable variable of physical interest bringing to light the
effectiveness and the wide potentiality of our method. In
particular we shall concentrate our attention on those vibrational
operators of clear physical meaning, for which the possibility of
applying the hypothesis of finite spectrum has already been
discussed in the previous section. Taking in mind the conclusions
reached in section II, we now therefore limit ourselves to show the
feasibility of the coupling mechanisms required in order to apply
our measurement procedure.

a) First of all let's consider the bosonic excitation number $\hat{n}_i$. In order to measure  the mean value of such an operator exploiting the method presented in this paper, the effective Hamiltonian model
\begin{equation}
  \hat{H}^{(I)}_I=\hbar\gamma\hat{n}_{i}\hat{\sigma}_{x}
\end{equation}
has to be implemented. In ref\cite{Vogel-Modelli} it is shown how
this model can be realised in laboratory.

b) Another important quantity to measure is the angular momentum
of the centre of mass of the confined particle. In
ref.\cite{Knight}, in connection with other purposes, it has been
implemented the interaction hamiltonian
\begin{equation}
  \hat{H}^{(I)}_I=\hbar\gamma\hat{L}_{i}\hat{\sigma}_{x}
\end{equation}
where $\hat{L}_{i}$ is the component of the orbital angular momentum along the direction $i=x,y,z$. Such an interaction is useful to reach our goal with respect to the angular momentum operator.

We remark that, at the best of our knowledge, our proposal provides for the first time a method exploitable for measuring this observable in trapped ion systems.

c) The next operator which we wish to apply our measurement protocol to, is a space-correlation operator defined as
\begin{equation}
   \hat{C}_{ij}\equiv\hat{a}_i\hat{a}^{\dag}_j+\hat{a}_j\hat{a}^{\dag}_i=
   \hat{Q}_i \hat{Q}_j + \hat{P}_i \hat{P}_j \;\;\;\;\;\;\; i\not=j
\end{equation}
with $i,j=x,y,z$ and where $\hat{Q}_k=\frac{1}{\sqrt{2}}(\hat{a}_k+\hat{a}_k^{\dag})$ and $\hat{P}_k=\frac{i}{\sqrt{2}}(\hat{a}_k^{\dag}-\hat{a}_k)$ are the \emph{dimensionless} ion center of mass position and momentum operators respectively.

The interaction hamiltonian model of the form (\ref{interaction})
corresponding to $\hat{A}=\hat{C}_{ij}$ may be easily implemented.
Consider for example two Raman schemes \cite{nist} responsible for
two couplings realised via two effective lasers directed along the
axis $x'$ and $y'$,  rotated by $\frac{\pi}{4}$ with respect to
$x$ and $y$ axis. Suppose that the two effective lasers relative
to each Raman scheme are tuned on the \emph{carrier} frequency
$\omega_A$, are $\pi$ out of phase and finally have wave vectors
of equal moduli. Under these hypotheses the total interaction
hamiltonian takes the form
\begin{equation}\label{C-medio-1}
  \hat{H}^{(I)}_I=\hbar\Gamma \left[
          f(\eta,\hat{a}_+^{\dag}\hat{a}_+)-f(\eta,\hat{a}_-^{\dag}\hat{a}_-)
                        \right] \hat{\sigma}_x
\end{equation}
where $\hat{a_{\pm}}=\frac{1}{\sqrt{2}}\left(\hat{a}_x \pm
\hat{a}_y\right)$, $\eta$ is the Lamb-Dicke parameter and $f$ an
assigned function giving rise to  non-linearity effects  stemming
from the non vanishing gradient of the electric
field\cite{nist,Vogel-Rass}.

If we restrict our analysis to the Lamb-Dicke limit (that is assuming $\eta<<1$), up to terms of order $\eta^2$, we just obtain
\begin{equation}\label{C-medio-2}
  \hat{H}^{(I)}_I=\hbar\gamma \hat{C}_{xy} \hat{\sigma}_x
\end{equation}
with $\gamma\propto\Gamma\eta^2$.

Measuring the operator $\hat{C}_{xy}$ is interesting, because, for
example, it provides a \textit{simple way for revealing the so
called \lq\lq Parity effect\rq\rq} predicted the first time in the
context of bimodal cavity QED\cite{Anna-Parity} and then in the context of
two-dimensional trapped ions\cite{Sabrina-Parity}. In ref.\cite{Sabrina-Parity} it has been proved that
a trapped ion, prepared in a $SU(2)$ coherent state and subjected
to the action of a suitable set of lasers whose effects are
described by a \emph{bimodal and two-boson Jaynes-Cummings
model},
$\hat{H}^{(I)}_I=\hbar\gamma\left(\hat{a}_x{a}_y\KetBra{+}{-}+h.c.\right)$,
exhibits a nonclassical behaviour that may be brought to light
measuring the mean value of $\hat{C}_{xy}$. This operator, indeed,
distinguishes in a \emph{macroscopic} way between two \emph{just
microscopically} distinguishable initial states.

More in detail, it has been analytically predicted that, if $N_0$
is the total number of bosonic excitations in the initial state
then, at particular time instants $t_p^*$ and $t_c^*$, it results
\begin{eqnarray}
  \MeanValue{\hat{C}_{xy}}(t_p^*)=
  \left\{
   \begin{array}{l}
     +N_0    \;\;\;\;\;\;\;\;\;\;\;\;  if \;\;\; N_0 \;\;\; odd \\
     -N_0    \;\;\;\;\;\;\;\;\;\;\;\;  if \;\;\; N_0 \;\;\; even
   \end{array}
  \right.
\end{eqnarray}
\begin{equation}
  \MeanValue{\hat{C}_{xy}^2}(t_c^*)=
  \left\{
   \begin{array}{l}
     +N_0^2    \;\;\;\;\;\;\;\;\;\;\;\; if \;\;\; N_0 \;\;\; odd \\
     -N_0^2    \;\;\;\;\;\;\;\;\;\;\;\; if \;\;\; N_0 \;\;\; even
   \end{array}
  \right.
\end{equation}
respectively.

The possibility of measuring the correlation $\hat{C}_{xy}$ and the operator $\hat{C}_{xy}^2$ is then useful to observe the \textit{\lq\lq Parity\rq\rq} discrimination.

d) As last example we consider the \emph{dimensionless} position operator
$\hat{Q}_x=\frac{1}{\sqrt{2}}(\hat{a}_x+\hat{a}_x^{\dag})\propto\hat{X}$.
Implementing the hamiltonian
\begin{equation}\label{Xmean}
  \hat{H}^{(I)}_I=\gamma(\hat{a}_x+\hat{a}_x^{\dag})\hat{\sigma}_x
\end{equation}
is very simple. Indeed it requires the use of two effective lasers both directed along $x$ and suitably tuned. One of the lasers must be first red sideband tuned and the other one must be tuned on the first blue sideband. With such a laser configuration one obtains
\begin{equation}
  \hat{H}^{(I)}_I=\gamma_1(\hat{a}_x\KetBra{+}{-}+\hat{a}_x^{\dag}\KetBra{-}{+})+
                  \gamma_2(\hat{a}_x^{\dag}\KetBra{+}{-}+\hat{a}_x\KetBra{-}{+})
\end{equation}

Adjusting the lasers intensities to have $\gamma_1=\gamma_2=\gamma$, the interaction hamiltonian model coincides with the one expressed by eq.(\ref{Xmean}).

Summarising, in this paper we have discussed a simple method aimed at measuring the mean value of vibrational observables. The key idea of our proposal is the possibility of transferring
information about the expectation value of a vibrational operator to the atomic populations, with the help of a suitable vibronic coupling. Applications discussed in this section highlight the relevance of the present procedure to a wide variety of situations.
Apart from some specific applications here considered, we emphasise that our method is also compatible with non-spherical traps. The reason is that our proposal rests upon the implementation of the interaction in eq.(\ref{interaction}), which does not depend on the trap isotropy.

Concluding we wish again to stress that, when we want to know only the expectation value of a physical quantity in an assigned vibrational state, the method presented in this paper might be easier to realise with respect to the others already quoted in the introduction.

\section*{Acknowledgements}
One of the authors (A. N.) acknowledges financial support from Finanziamento Progetto Giovani Ricercatori anno 1998, Comitato 02.

\end{document}